\documentclass[sn-mathphys-ay]{sn-jnl}

\usepackage{graphicx}%
\usepackage{multirow}%
\usepackage{amsmath,amssymb,amsfonts}%
\usepackage{amsthm}%
\usepackage{mathrsfs}%
\usepackage[title]{appendix}%
\usepackage{xcolor}%
\usepackage{textcomp}%
\usepackage{manyfoot}%
\usepackage{booktabs}%
\usepackage{algorithm}%
\usepackage[algo2e]{algorithm2e} 
\usepackage{algorithmicx}%
\usepackage{algpseudocode}%
\usepackage{listings}%
\usepackage{tikz}%
\usepackage{lmodern}%
\usepackage{multicol}

\begin{document}

\title[Article Title]{
Unveiling land use dynamics: Insights from a hierarchical Bayesian spatio-temporal modelling of Compositional Data}

\author*[1]{\fnm{Mario} \sur{Figueira}}\email{Mario.Figueira@uv.es}
\author[1]{\fnm{Carmen} \sur{Guarner}}
\author[1]{\fnm{David} \sur{Conesa}}
\author[1]{\fnm{Antonio} \sur{López-Quílez}}
\author[2]{\fnm{Tamás} \sur{Krisztin}}

\affil*[1]{\orgdiv{Department of Statistics and Operations Research}, \orgname{University of Valencia}, \orgaddress{Carrer del Dr. Moliner, 50, 46100 Burjassot, Valencia, Spain}}

\affil[2]{\orgname{International Institute for Applied Systems Analysis (IIASA)}, \orgaddress{Schlossplatz 1, 2361, Laxenburg, Austria}}



\abstract{Changes in land use patterns have significant environmental and socioeconomic impacts, making it crucial for policymakers to understand their causes and consequences. This study, part of the European LAMASUS (Land Management for Sustainability) project, aims to support the EU's climate neutrality target by developing a governance model through collaboration between policymakers, land users, and researchers. We present a methodological synthesis for treating land use data using a Bayesian approach within spatial and spatio-temporal modeling frameworks.

The study tackles the challenges of analyzing land use changes, particularly the presence of zero values and computational issues with large datasets. It introduces joint model structures to address zeros and employs sequential inference and consensus methods for Big Data problems. Spatial downscaling models approximate smaller scales from aggregated data, circumventing high-resolution data complications.

We explore Beta regression and Compositional Data Analysis (CoDa) for land use data, review relevant spatial and spatio-temporal models, and present strategies for handling zeros. The paper demonstrates the implementation of key models, downscaling techniques, and solutions to Big Data challenges with examples from simulated data and the LAMASUS project, providing a comprehensive framework for understanding and managing land use changes.}


\keywords{Land use, compositional data, spatio-temporal models, downscaling, Big Data, INLA}

\maketitle

\section{Introduction}\label{sec:Introduction}


Changes in land use patterns have significant environmental and socioeconomic impacts, making it crucial for policymakers to understand their causes and consequences \citep{LeSage_IntroductionSpatialEconometrics_2008, Chakir_PredictingSpatialLandUse_2013}. Studies on land use are valuable for identifying the determinants of these changes, which affect biodiversity, water pollution, soil erosion, climate change, and economic and social welfare \citep{Hersperger_LandscapeChangeCentralEurope_2009, Chakir_AgriculturalRentLandUse_2017, Moindjie_InteractionsLandUse_2022, Bareille_StructuralIdentificationCropYields_2024}. These changes are driven by a combination of socioeconomic factors, pedo-climatic conditions, and policy variables. Land use models are crucial for analysing these influences and their effects \citep{Chakir_DeterminantsLandUse_2009, Hersperger_LandscapeChangeCentralEurope_2009, vanVliet_DriversAgriculturalLandUse_2015}. Moreover, the decision-making processes surrounding land use are complex, shaped by both local and global biophysical and socioeconomic factors. Therefore, comprehensive knowledge of these influences aids in evaluating and formulating environmentally friendly public policies.


Various disciplines, including economics, statistics, geography, and land use science, have developed empirical land use modelling approaches using aggregate or individual data \citep{Chakir_SpatialDownscaling_2009}. However, many studies overlook spatial autocorrelation in modelling land use or use ad hoc methods \citep{LeSage_IntroductionSpatialEconometrics_2008, Chakir_PredictingSpatialLandUse_2013, Chakir_SpatialAutocorrelation_2021, Moindjie_InteractionsLandUse_2022}, despite its prevalence in economic decisions. This underscores the need for more sophisticated spatial econometric models that account for spatial heterogeneity and interdependence to accurately measure and analyse land use patterns \citep{LeSage_IntroductionSpatialEconometrics_2008, LeSage_IntroductionSpatialEconometrics_2009, Elhorst_SpatialEconometrics_2013, Chakir_SpatialAutocorrelation_2021}.


This work is part of the European \href{https://www.lamasus.eu/}{LAMASUS} (Land Management for Sustainability) project. LAMASUS develops an innovative governance model through collaboration between policymakers, land users, and researchers. This Horizon Europe project aims to support the European Union’s climate neutrality target by creating an open-access modelling toolbox for designing effective land use policies within the framework of the European Green Deal. In the context of the LAMASUS project, this paper presents a methodological synthesis for the treatment of land use data using a Bayesian approach and in the spatial and spatio-temporal modelling framework. This approach is integrated in conjunction with those that aim to provide a better understanding of the spatial dynamics of the processes that determine and drive land use changes. Enhanced understanding also enables the implementation of land use policies based on models that improve policy formulation.

LAMASUS faces different problems related to land use and land use change analysis. In this case, there are numerous land use categories, which are aggregated into a small set to facilitate the determination of the main drivers of land use at different scales, both administrative (NUTS scales) and high-resolution levels, that is, small spatial scales. From the aggregated spatial scales, to approximate a smaller scale, spatial downscaling models or disaggregation models are applied, avoiding dealing with high-resolution data that may present inconveniences related to the systematic presence of 0's or being large data banks, which would pose a Big Data problem. On the other hand, to work with the aggregated data with many disaggregated land use categories, joint model structures are proposed to deal with the systematic presence of 0's and 1's, while for problems that emerge from Big Data, a combined procedure of sequential inference and consensus is proposed.



In particular, the aim of this paper is to present different procedures and methods to analyse land use data in different contexts of spatial and spatio-temporal modelling. In this sense, section 2 introduces land use data and its analysis by Beta regression, when only one land use category is available, or within the framework of Compositional Data Analysis (CoDa), if multiple land use categories are available. In Section 3 we briefly review a selection of those spatial and spatio-temporal models that are of interest for assessing the spatial and spatio-temporal structure of land use data. Section 4 sets out the proposed methodology for dealing with the presence of 0's and 1's in land use data, both when a single category is available and when several categories are available. In Section 4 is briefly shown the implementation of the main spatial and spatio-temporal models for land use data. Section 5 illustrates how to deal with the presence of 0's and 1's in Beta regression and CoDa analysis. In Section 6 we present downscaling models, also known as disaggregation models, applied to land use and compositional data. Section 7 is related to Big Data in the framework of high-resolution land use data. Section 8 stands for some simulated data and real data from LAMASUS project. Finally, we conclude in Section 9.

\section{Land use and compositional data}

In this section, we focus on presenting the treatment of compositional data for the analysis of land use shares, providing a brief exposition of the theoretical framework for CoDa and the implementation of such analysis to address the drivers and underlying process in land use shares. 


Land use data is usually presented as proportions of land use shares in a given area, which implies that land use shares, by definition, have a compositional nature \citep{Pirzamanbein_LandCoverCompositional_2020, ThomasAgnan_SpatialSimultaneousAutoregressive_2021, Krisztin_SpatialMultinomialLandUse_2022}. Therefore, one approach to capture the variability of land use shares would be through Beta regression if we focus on a specific category, or through CoDa to analyse the joint variability of several land use categories, e.g. cropland, grassland, forest, urban and other. The joint analysis of the categories in the composition can be considered by different approaches: transforming the logratios \citep{Aitchison_CompositionalData_1986, Aitchison_CompositionalData_2005, Greenacre_AitchisonCompositionalReappraisal_2023}, and modelling those logratios using multivariate Gaussian models, or using Dirichlet regression models \citep{Connor_DirichletDistribution_1969, Hijazi_DirichletData_2009} for the proportions without needing to transform them.





In general, compositional data consists in a set of \textit{parts} that identifies the constituents of the composition $\{1, 2, \ldots, D\}$, while \textit{components} $\{\mathbf{y}_{1}, \mathbf{y}_{2}, ..., \mathbf{y}_{D}\}$ are numerical proportions in which individual parts occur \citep{Aitchison_CompositionalData_1986}. Therefore, CoDa can be defined by a matrix $\mathbf{Y}_{n \times D}$ of $n$ observations times $D$ parts of the composition. Each row of the matrix $\mathbf{Y}_{n \times D}$ satisfies the following closure condition $\sum_{d=1}^D \mathbf{Y}_{id}=1$, and for each component is usually to assume for each value that $\mathbf{Y}_{id} \in (0,1)$. The vector $\mathbf{y}_{1\times D}$ for the $i$-th observation, related to each row of $\mathbf{Y}_{n \times D}$, is called composition and it pertains to the simplex sample space. A simplex space with dimension $d$, denoted by $\mathbb{S}^d$, is defined as:

\begin{equation}
    \mathbb{S}^D = \left\lbrace \mathbf{y} \in \mathbb{R}^D : y_{d}\in (0,1), \sum_{d=1}^D y_d = 1 \right\rbrace,
\end{equation}
where due to the closure condition and the property $y_i \in (0,1)$, the entire set of observations of the composition $\mathbf{Y}_{n \times D}$  does not follow the usual Euclidean geometry $\mathbb{R}^D$, but rather the Aitchison geometry $\mathbb{S}^d$ with dimension $d = D - 1$ \citep{Aitchison_CompositionalData_1986, Greenacre_AitchisonCompositionalReappraisal_2023}. Additionally, it is also possible to defined a \textit{subcomposition} $\mathbf{y}_{1 \times C}$ of $\mathbf{y}_{1 \times D}$. This refers to a subset of $C$ parts of the referring composition of $D$ parts, where $C < D$ and the subcomposition is also subject to the closure condition $\sum_{i=1}^C y_{i} = 1$.




\subsection{Logratio transformations}

In Aitchison's approach to compositional data analysis \citep{Aitchison_CompositionalData_1982, Aitchison_CompositionalData_1986}, the focus was on distributional issues, such as finding a way to transform compositions into interval-scale multivariate vectors that could validly use the multivariate normal distribution. The proposed logistic transformations, which aimed to create ``transformed-normal'' models, relied on logarithmic transformations. These transformations allowed the use of standard unconstrained multivariate statistics applied to transformed data, with inferences translatable back into compositional statements \citep{Aitchison_CompositionalData_1986, Greenacre_AitchisonCompositionalReappraisal_2023}. However, this reliance on logarithms precluded zero data values, leading to the ongoing debate about zero replacement and treatment, a topic often overlooked in publications \citep{Greenacre_AitchisonCompositionalReappraisal_2023}.

There are several proposals for logratio transformations to analyse CoDa, including those from Aitchison's original work \citep{Aitchison_CompositionalData_1982, Aitchison_CompositionalData_1986}, such as \textit{pairwise logratios} (LR), \textit{additive logratios} (ALR), and \textit{centred logratios} (CLR). Later proposals \citep{Egozcue_IsometricLogratio_2003, Egozcue_SimplicialGeomtryCoDa_2006, Fiserova_OrthonormalCoordinatesCoDa_2011} include \textit{isometric logratios} (ILR) and \textit{pivot logratios} (PLR). These transformations attempt to address various shortcomings of the alternative methods. However, each logratio transformation has its own strengths and weaknesses, making it impossible to propose a single transformation suitable for all CoDa datasets \citep{Greenacre_IsometricLogratioEvaluation_2019, Greenacre_AitchisonCompositionalReappraisal_2023}. 

In this work, we will focus on two logratios of the proposed by \citep{Aitchison_CompositionalData_1986}: ALR and CLR. The ALR transformation implies defining a subset of $d=D-1$ logratios with the same denominator, called the \textit{reference} ($r$) part. Therefore, the ALR with respect to a reference part of the composition is written as 
\begin{equation}
    \text{ALR}(y_i \mid y_r) = \log\left(\frac{y_i}{y_r}\right), \ i \in \{1,..., D\}:i\neq r,
\end{equation}
where is possible to define $D$ alternative ALR transformations, depending on the part chosen as reference. In this case, it is important to use a reference part that satisfy either statistical or meaningful objective. In fact, if the chosen reference part is almost constant, then the corresponding ALRs are approximately the logarithm of the parts, up to a nearly constant amount.





The CLR transformation is defined as the logratios of the parts with respect to their geometric mean $g(\mathbf{y})=\prod_{i=1}^Dy_i^{1/D}$:
\begin{equation}
    \text{CLR}(y_i) = \log\left(\frac{y_i}{g(\mathbf{y})}\right), \ i \in \{1,...,D\},
\end{equation}
transformation that turns the sum-to-one constraint into a sum-to-zero constraint $\sum_{i=1}^D \text{CLR}(y_i) = \sum_{i=1}^D \left[log(y_i) - \log(g(\mathbf{y})) \right] = 0$. The usefulness of this transformation is that Euclidean distances between the CLRs are identical to the distances using all possible logratios; in other words, the CLR is an isometric transformation.

Both logratio transformations allow us to model the logratios using multivariate Gaussian distributions $\mathbf{y}^* \sim \text{MVN}(\boldsymbol\mu, \boldsymbol\Sigma)$, where $\mathbf{y}^*$ represents the logratios, either from CLR or ALR transformations. This multivariate model not only accounts for complex structures in the linear predictors, including spatial and spatio-temporal dependencies, but also captures the correlations between the logratios through the variance-covariance matrix $\boldsymbol\Sigma$ of the multivariate Gaussian distribution \citep{Minaya_CoDaINLA_2024}.



\subsection{Dirichlet distribution}

Dirichlet distribution is a generalisation of the Beta distribution for more than two proportions, and its probability density function is defined as 
\begin{equation}
    \pi(\mathbf{y}\mid \boldsymbol\alpha) = \frac{1}{\text{B}(\boldsymbol\alpha)} \prod_{i=1}^D y_i^{\alpha_i - 1},
\end{equation}
where $\boldsymbol\alpha = (\alpha_1,...,\alpha_D)$ is the vector of shape parameters $\alpha_i>0 \ \forall i$, and $\text{B}(\boldsymbol\alpha)$ is the Multinomial Beta function, being the normalizing constant of the Dirichlet. The sum of the shape parameters $\alpha_0 = \sum_{i=1}^D\alpha_i$ is usually interpreted as a precision marginal parameter $\tau$. In the Dirichlet distribution the sum-to-one constraint $\sum_{i=1}y_i = 1$ is also satisfied by the data vector $\mathbf{y} = (y_1,..., y_D)$, $y_i \in (0,1)$. Therefore, if $\mathbf{y} \sim \mathcal{D}(\boldsymbol\alpha)$ denote a variable that is Dirichlet distributed, the expected values are $\mathbb{E}(y_i) = \alpha_i/\alpha_0$, and the relation between the shape parameters and the linear predictor is given by the log-link function $\log(\alpha_i)=\eta_i$. Thus, it is also possible to parameterise the Dirichlet distribution in terms of the mean as $\mu_{i} = \exp(\eta_{i})/(\sum_{i=1}^D \exp(\eta_{i}))$.



The parameterization of the Dirichlet distribution in terms of the shape parameters (or the mean) linked to the linear predictor allows us to use complex structures, as in the INLA framework, by defining Latent Gaussian Models (LGM). These structures can include temporal components, spatial components such as the SPDE-FEM approach, or more complex spatio-temporal processes. The implementation of the Dirichlet distribution in \texttt{R-INLA} relies on methodology and R-package presented by \citep{Minaya_DirichletINLA_2023}. This provides the basis to implement complex spatial and spatio-temporal structures for Dirichlet distributed data inside the INLA framework.




\section{Spatial and spatio-temporal models}

The land use may present changes along space and time. Consequently, to assess the variability it is possible to construct spatio-temporal models that can provide a further description and insights \citep{LeSage_IntroductionSpatialEconometrics_2008, Chakir_SpatialAutocorrelation_2021}. In this section, we will provide a brief overview of the different structures that can be used in spatio-temporal modelling for land use data.



In most cases, land use data is provided as areas, i.e., values related to polygons that typically represent administrative boundaries. This areal data naturally arises when a fixed domain is partitioned into a finite number of subregions where outcomes are aggregated. Therefore, an appropriate way to model this kind of spatial dependency is by building a structure of adjacency, or an adjacency matrix $\mathbf{W}$, between the different connected areas and setting a multivariate Gaussian density with a precision matrix based on a modulation of this adjacency structure \citep{Banerjee_HierarchicalSpatialData_2015}, which is quite common in disease mapping as well \citep{Beneito_DiseaseMapping_2019}. This prior distribution defines the dependence between the connected areas and allows the assessment of spatial structures, from simpler ones like the Besag model \citep{Besag_SpatialModelBesag_1974} or the Besag-York-Mollié model \citep{Besag_SpatialModelBYM_1991} to more complex multidimensional spatial structures \citep{Beneito_MModels_2017}. 



In spatial econometrics, the Spatial Error model (SEM), Spatial Lag model (SLM), and Spatial Durbin model (SDM) are commonly used to analyse spatial dependency \citep{Elhorst_SpatialEconometrics_2013, Bivand_SpatialEconometricsINLA_2014}. These models also define a spatial dependence structure for the explanatory covariates and the error term. The general framework that synthesises these models is the general nesting spatial model equation \citep{Elhorst_SpatialEconometrics_2013}
\begin{equation}
\begin{array}{c}
     \mathbf{y} = \delta \mathbf{W} \mathbf{y} + \alpha \mathbf{1}_{N} + \mathbf{X}\boldsymbol\beta + \mathbf{W}\mathbf{X}\boldsymbol\theta + \mathbf{u}, \\
     \mathbf{u} = \lambda \mathbf{W} \mathbf{u} + \varepsilon,
\end{array}
\end{equation}
where $\mathbf{1}_N$ denotes a vector ($N \times 1$) of ones, $\alpha$ is a global intercept, $\mathbf{W}$ s an adjacency matrix of positive defined weights usually normalized such that each row sums to unity \citep{LeSage_IntroductionSpatialEconometrics_2008, LeSage_IntroductionSpatialEconometrics_2009}, $\mathbf{W}\mathbf{y}$ represents the endogenous interaction effects of the response variables $\mathbf{y}$, $\mathbf{W}\mathbf{X}$ represents the exogenous interaction effects among the explanatory variables $\mathbf{X}$, $\mathbf{W}\mathbf{u}$ denotes the interaction among the spatial units, $\boldsymbol\beta$ and $\boldsymbol\theta$ represent unknown parameters. Finally, $\delta$ and $\lambda$ are the spatial autoregressive parameter and spatial autocorrelation parameter, respectively. These models can be reassembled to fit within the INLA framework \citep{Bivand_SpatialEconometricsINLA_2014, Virgilio_SpatialEconometricsINLA_2021} by fixing some parameters and performing their analysis outside INLA, using Monte Carlo (MC, \citealt{Berild_INLAandMC_2022}) or Markov Chain Monte Carlo (MCMC, \citealt{Virgilio_INLAandMCMC_2018}).


The temporal terms can be incorporated into the linear predictor of the model in an additive way, i.e. as a new term evaluating a purely temporal trend. This can be synthesised into a model with one spatial component and one temporal component $\mathbf{y} = \mathbf{u}_s + \mathbf{u}_t + \varepsilon$, where $\mathbf{u}_s$ is the purely spatial component, $\mathbf{u}_t$ is the purely temporal term and $\varepsilon$ is an independent and identically distributed error term. As an example, the temporal component can be integrated into a rewritten form of the SLM to fit within the INLA framework in the following way:
\begin{equation}
    \mathbf{y} = (\mathbf{I} + \rho\mathbf{W})^{-1} \mathbf{X} \boldsymbol\beta + \mathbf{u}_t + \mathbf{u}_s
\end{equation}
where $u_{t,i} = u_{t,i-1} + \epsilon(\sigma)$ follows a first-order random walk, and $\mathbf{u}_s \sim \text{MVN}(\mathbf{0}, \mathbf{Q}):\mathbf{Q} = \tau (\mathbf{I}-\rho_l\mathbf{W})^T \cdot (\mathbf{I}-\rho_l\mathbf{W})$ represents the rewritten spatial structure for the spatial error component.


In addition to being able to evaluate additive spatio-temporal models, it is possible to consider different structures for spatial and temporal interaction components. In \citep{KnorrHeld_SpaceTime_2000} four general structures are proposed to define spatio-temporal interaction models, in which the precision matrix of the spatio-temporal effect is constructed by means of the Kronecker product of the precision matrix of the spatial component by the precision matrix of the temporal component $\mathbf{Q}_{st}=\mathbf{Q}_s\otimes\mathbf{Q}_t$ \citep{Clayton_GLMM_1996}. Four natural interactions arise from this formulation: (a) Type I interaction, this is defined as the Kronecker product of the precision matrices of an unstructured spatial $\mathbf{Q}_s=\tau_s\mathbf{I}$ and temporal effect $\mathbf{Q}_t=\tau_t\mathbf{I}$, where the structure for the spatio-temporal component is $\mathbf{Q}_{st}=\mathbf{Q}_s\otimes\mathbf{Q}_t=\tau_{st}\mathbf{I}\otimes\mathbf{I}$. This interaction makes sense when space and time have a discretised structure, otherwise this specification may not make sense. (b) Type II interaction assumes an unstructured spatial effect $\mathbf{Q}_s=\tau_s\mathbf{I}$ interacting with a structured spatial effect $\mathbf{Q}_t$, then the spatio-temporal precision matrix is $\mathbf{Q}_{st}=\mathbf{I}\otimes\mathbf{Q}_t$. (c) Type III interaction is the interaction of a structured spatial effect $\mathbf{Q}_s$ by an unstructured temporal effect $\mathbf{Q}_t=\tau_t\mathbf{I}$, obtaining that $\mathbf{Q}_{st}=\mathbf{Q}_s\otimes\mathbf{I}$. (d) Type IV interaction is the interaction between a structured spatial effect $\mathbf{Q}_s$ and a structured temporal effect $\mathbf{Q}_t$, which implies that the interacting spatio-temporal component is defined by $\mathbf{Q}_{st}=\mathbf{Q}_s\otimes\mathbf{Q}_t$. Moreover, it is also possible to construct non-separable spatio-temporal interaction models \citep{Prates_NonSeparableSpatioTemporal_2022, Lindgren_diffusionbased_2024}. In these models, the precision matrix cannot be decomposed into the Kronecker product of a spatial precision matrix and a temporal precision matrix, i.e., $\mathbf{Q}_{st} \neq \mathbf{Q}_s \otimes \mathbf{Q}_t$.

\section{Implementing spatial CoDa models}

Implementing spatial models in compositional data can be challenging due to the multivariate nature of the response variables. Each category or log-ratio of land use may have its own spatial or spatio-temporal component with a distinct set of hyperparameters. To address this complexity, rather than defining a separate component for each variable, two alternatives can be considered. One approach involves replicating some spatial or spatio-temporal components across different land use categories or log-ratios, resulting in identical hyperparameters for these replicated components \citep{Virgilio_BayesianRINLA_2020}. Another approach is to share spatial or spatio-temporal components across some categories, either scaled by parameters to be estimated or sharing the same value \citep{Virgilio_BayesianRINLA_2020}. 

This can be exemplified with the following toy model for three logratios $\mathbf{Y} = (\mathbf{y}_1, \mathbf{y}_2, \mathbf{y}_3)$, where $\mathbf{Y} \sim \text{MVN}(\boldsymbol\mu, \boldsymbol{\Sigma})$, and the mean vector $\boldsymbol{\mu}=(\boldsymbol\mu_1, \boldsymbol\mu_2, \boldsymbol\mu_3)$ is modelled as:
\begin{equation}
\begin{array}{c}
    \boldsymbol\mu_{1} = \beta_{10}\mathbf{1} + \mathbf{X}\boldsymbol\beta_{11} + \mathbf{u}_{1s}(\tau_s, \lambda_s) + \mathbf{u}_{t}(\tau_t), \\
    \boldsymbol\mu_{2} = \beta_{20}\mathbf{1} + \mathbf{X}\boldsymbol\beta_{21} + \mathbf{u}_{2s}(\tau_s, \lambda_s) + \mathbf{u}_{t}(\tau_t), \\
    \boldsymbol\mu_{3} = \beta_{30}\mathbf{1} + \mathbf{X}\boldsymbol\beta_{31} + \mathbf{u}_{3s}(\tau_s, \lambda_s) + \mathbf{u}_{t}(\tau_t), \\
\end{array}
\end{equation}
where $\beta_{\bullet0}$ are the intercepts, $\boldsymbol\beta_{\bullet1}$ represents the fixed effect vector related to the matrix of explanatory variables $\mathbf{X}$, $\mathbf{u}_{\bullet s}$ denotes the spatial effects, which are three different replicas from the same Leroux distribution \citep{Leroux_NewSpatialModelDependence_1999}, given by $\mathbf{u} \sim \text{MVN}(\mathbf{0}, \mathbf{Q}_s):\mathbf{Q}_s = \tau_s \cdot [\mathbf{I} + \lambda_s \cdot (\mathbf{D} - \mathbf{I} - \mathbf{W})]$, and $\mathbf{u}_t: u_{t,i} = u_{t,i-1} + \varepsilon_i(\tau_t)$ is a temporal component shared through the three linear predictors with the same values for each logratio. Finally, $\boldsymbol\Sigma$ is the variance-covariance matrix between of logratios, with $\boldsymbol\Sigma^{-1}_{ii} = \tau_{i}$ and $\boldsymbol\Sigma^{-1}_{ij} = \rho_{ij}/(\tau_i \cdot \tau_j)^{1/2} $, $i \neq j$. 

In \texttt{R-INLA} it is not possible to model multivariate models directly, so the correlation structure for the multivariate normal observations can be represented by a random effect with such a structure in the linear predictor of the model. To illustrate this, suppose we have $n$ observations of a $D$-dimensional multivariate distribution. To evaluate the correlations of this multivariate distribution in \texttt{R-INLA}, we can construct a component of the latent field $\mathbf{u}$ such that $\mathbf{u} \sim \text{MVN}(\mathbf{0}, \mathbf{Q})$, whose precision matrix is defined as $\mathbf{Q}_{ii}=\tau_{ii}$ and $\mathbf{Q}_{ij}= \rho_{ij}/\sqrt{\tau_{ii}\tau_{jj}}$. Each realisation of this component produces a vector of $D$ elements, so if its realisation is replicated for each observation, we will have $n$ replicas of this effect evaluated with the same hyperparameters $\{\boldsymbol\tau, \boldsymbol\rho\}$. This allows us to express the $D$-dimensional multivariate distribution with $n$ components as a univariate distribution with $n \times D$ observations and fixed precision in the univariate Gaussian likelihoods to a high value ($\log(\tau)\gg 1$), where in each term of the linear predictor we will have a realisation from this effect $u$ corresponding to a category $d \in \{1,...,D\}$ and an observation $i \in \{1,...,n\}$.


This demonstrates the flexibility to define various spatio-temporal structures for modelling compositional data, or more generally, multivariate data. It also highlights the combinatorial challenge of defining models within such a framework, where testing all possible combinations implies defining $2^D$ models for each component of the model, with $D$ being the multivariate dimensionality. This issue suggests that constructing models in these contexts can benefit significantly from expert knowledge, which can help narrow down the feasible set of reasonable models. This approach avoids the need for extensive model selection procedures among numerous sets, which can become intractable with even a few components. Therefore, implementing spatial models in the context of compositional data requires establishing criteria based on expert knowledge or other relevant factors to limit the set of models to be evaluated. This is especially important given the various possibilities for defining spatio-temporal components that we have mentioned in the previous section.

\section{Dealing with 0's in land use data}

In the first section, we presented various approaches to analyse land use data: beta models when focusing on a single category, and multivariate normal models for some transformation of logratios (additive logratios, centred logratios, isometric logratios) or Dirichlet models when dealing with multiple categories. These are all different approaches to CoDa analysis in general, and to land use data provided in proportions in particular. However, a significant challenge in dealing with compositional data across all these approaches is the presence of zeros and ones \citep{Martin_RoundedZeroesCoDa_2006, Martin_DealingWithZerosCoDa_2011, Tsagris_DirichletCoDaZeros_2018}. Zeros can be problematic because they often indicate the absence of a component, complicating the application of standard statistical techniques that assume strictly positive data. Similarly, ones can signify the complete dominance of a single component (inducing zero values in the remaining components), which can distort correlations and overall analysis. Addressing these issues typically requires specific methods, such as zero-replacement strategies or transformations, to ensure that the compositional nature of the data is properly accounted for and that the analyses yield meaningful results.



In general, different types of zeros can be found in compositional data (CoDa): (i) rounded zeros or below-detection values, (ii) count zeros, and (iii) essential zeros. The first two types of zeros are associated with ``false'' zeros, where the structure of the experimental design, the sample size, or the sensitivity of the measuring instruments results in null values, even though the underlying phenomenon is not necessarily null. In contrast, the third type of zeros, also called structural or absolute zeros, assumes that the measurements indeed capture the real absence or null value of the phenomenon \citep{Martin_DealingWithZerosCoDa_2011}. For the first two cases of rounded zeros and count zeros, replacement procedures or transformations have been proposed \citep{Rasmussen_ZerosCoDa_2020, Lubbe_CoDaZeroReplacement_2021}, while for essential zeros, likelihood modifications such as zero-inflated models or hierarchical structures can be used \citep{Aitchison_PossibleSolutionsZeros_2003, Tsagris_DirichletCoDaZeros_2018, Tang_DirichletZeros_2022}.

In line with \citep{Aitchison_PossibleSolutionsZeros_2003}, a joint hierarchical model similar to Hurdle models \citep{Mullahy_CountDataHurdle_1986, Cameron_PoissonRegressionHurdle_2013, MartinezMinaya_SPDsSpatioTemporal_2018}, which are widely used in the environmental and ecological sciences, can be proposed to deal with null values in likelihoods that do not allow them to be evaluated. This approach involves conditional modelling of the components according to whether each particular component has a zero or non-zero value. In order to apply a structure similar to that of a Hurdle model, there must be a correspondence in the modelling of the composition and subcompositions with that of the null and non-null values of the subcompositions. This implies that we can use it in the case of dealing with data distributed according to a Dirichlet or when we use the centred logratios, since it allows us to make the joint analysis of the zeros with the process that generates the logratios for the CLR, or the shape parameters in the case of the Dirichlet.




Let's assume that we have a matrix of $n$ observations and $D$ compositions, $\mathbf{Y}_{n \times D}$ where each element $Y_{ij}$ is the value for the $i$-th observation and $j$-th composition. Then, we can define an incidence matrix $\mathbf{I}_{n \times D}$, that for each element $I_{ij}$ we have $I_{ij} = 0 \iff Y_{ij} = 0$ and $I_{ij} = 1 \iff Y_{ij} \neq 0$. Each column of the incidence matrix will be the values that we will model according to a Bernoulli distribution, and in principle this modelling will be independent between the different compositions. Thus, for each composition, the vector $\mathbf{i}_d = \mathbf{I}_{1:n,d}$ we will have:
\begin{equation}
\begin{array}{rcl}
     \mathbf{i}_{1} & \sim & \text{Ber}(\boldsymbol\pi_{1}) \\ 
     \text{logit}(\boldsymbol\pi_{1}) & = & \mathbf{X}\boldsymbol\beta_1 + \sum_{k=1}^K f_{k,1}(\mathbf{z}_k) + \mathbf{u}_{st,1} \\ 
     & \vdots & \\
     \mathbf{i}_{D} & \sim & \text{Ber}(\boldsymbol\pi_{D}) \\
     \text{logit}(\boldsymbol\pi_{D}) & = & \mathbf{X}\boldsymbol\beta_D + \sum_{k=1}^K f_{k,D}(\mathbf{z}_k) + \mathbf{u}_{st,D} \\ 
\end{array}
\end{equation}
where $\boldsymbol\beta_d$ is the vector of linear coefficients, $\mathbf{f}_d$ are non-linear components and $\mathbf{u}_{st,d}$ the spatio-temporal components for the $d$-th component. While for the models for the values of the compositions, if we use the CLR we can write $\mathbf{Y}^*=\text{CLR}(\mathbf{Y})$, we have the following multivariate Gaussian model for the logratio of the compositions $\mathbf{Y}^* \sim \text{MVN}(\mathbf{0},\boldsymbol\Sigma)$. However, in this case, it is important to distinguish between rows where there is a component with a zero value, or where there is one component with a unit value and the rest are zeros, and rows where no values are zero, in order to analyse them analogously to how it is done with Hurdle models.









The multivariate model can be rewritten as $D$ univariate Gaussian models in which the marginal precision is fixed at a very high value and the variance-covariance structure is evaluated by a random effect as described in the previous section in order to implement these models in \texttt{R-INLA}:
\begin{equation}
\begin{array}{rcl}
     \mathbf{y}^*_{1} & \sim & \text{N}(\boldsymbol\mu_{1}, \tau^*) \\ 
     \boldsymbol\mu_1 & = & \mathbf{X}\boldsymbol\beta_1 + \sum_{k=1}^K f_{k,1}(\mathbf{z}_k) + \mathbf{u}_{st,1} + \mathbf{u} \\ 
     & \vdots & \\
     \mathbf{y}^*_{D} & \sim & \text{N}(\boldsymbol\mu_{D}, \tau^*) \\ 
     \boldsymbol\mu_D & = & \mathbf{X}\boldsymbol\beta_D + \sum_{k=1}^K f_{k,D}(\mathbf{z}_k) + \mathbf{u}_{st,D} + \mathbf{u} \\ 
\end{array}
\end{equation}
where $\tau^*$ is a fixed precision with a high value ($\log(\tau^*) \gg 1$), $\mathbf{u}$ is the random effect that accounts for the variance-covariance structure, $\mathbf{u} \sim \text{MVN}(\mathbf{0}, \mathbf{Q})$, such that $\mathbf{Q}=\boldsymbol\Sigma^{-1}$. In addition, for each observation row with a zero value component, the CLR transformation is re-evaluated to exclude those null values. This procedure is the same whether we use the Dirichlet distribution or the Beta distribution, which is a simplification of the Dirichlet when only two categories are available.

\section{Downscaling models}

Empirical studies involving land use change often use aggregate data for regions, countries, or other geographic scales \citep{Chakir_SpatialDownscaling_2009, Chakir_PredictingSpatialLandUse_2013}. However, some land use studies have had access to individual parcel-level data for analysing water quality, the extent of urban sprawl, carbon sequestration costs, and habitat fragmentation \citep{Liangzhi_SpatialDisaggregationEntropy_2006, Chakir_SpatialDownscaling_2009}. The spatial pattern of land use is of particular interest and has been useful in identifying factors that drive changes in land use at the disaggregated level \citep{Chakir_SpatialDownscaling_2009, An_SpatialDisaggregationLandUse_2020}. In a more general sense, this issue of aggregated, disaggregated, and different scales of data availability and analysis can be related to fusion models \citep{Wang_DataFusion_2018, Wang_FusionModelUnifyingFramework_2021, Villejo_FusionModelINLA_2023}, data misalignment \citep{Moraga_SpatialMisalignment_2017}, change of support \citep{Gelfand_SpatialChangeOfSupport_2001, Bradley_SpatialChangeOfSupport_2016} and disaggregation modelling or downscaling \citep{Nandi_disaggregationRPackage_2023}. In this general context, different issues arise concerning the availability at different scales of information concerning the response variable, as in data fusion models \citep{Wang_FusionModelUnifyingFramework_2021}, or availability at different scales of relative information, as in downscaler models \citep{Liangzhi_SpatialDisaggregationEntropy_2006, Berrocal_SpatioTemporalDownscaler_2010}. This implies extracting information from an aggregated scale to a disaggregated or continuous scale using an effect with a continuous spatial or spatio-temporal structure, structure that is commonly used in geostatistics to account for the data spatial dependence. This is possible to implement in INLA thanks to the definition of spatial models using the SPDE-FEM approach, which is integrated in the \texttt{R-INLA} package \citep{Lindgren_SpatialINLA_2015, Bakka_SpatialINLAReview_2018}.

\subsection{SPDE-FEM}

The SPDE-FEM approach (which stands for \textit{Stochastic Partial Differential Equations} and \textit{Finite Element Methods}) allows for expressing the continuous spatial structure by approximating a Gaussian Field with a Gaussian Markov Random Field. This makes the inferential process efficient by reducing computational costs. This approach was initially proposed in \citep{Lindgren_ExplicitLinkSPDE_2011} by means of an approximate stochastic weak solution to the SPDE in Equation (\ref{eq:SPDEformulaWeakSolution}), proving that a stationary SPDE solution has a Matérn covariance function and allowing to calculate directly its precision matrix, bypassing the need for resource-intensive inverse processes. Thus, formally we get that the structured spatial effect $\xi$ with precision $\tau$ is related to a non-structured Gaussian random effect (white noise) through a filtering operator $(\kappa - \Delta)^{\alpha/2}$: 
\begin{equation}
\begin{array}{c}
    (\kappa^2-\Delta)^{\alpha/2}\tau \cdot \xi(\mathbf{s})=\mathcal{W}(\mathbf{s}),  \\
    \mathbf{s}\in \mathcal{D}, \ \alpha=\nu+ d/2, \ \kappa>0, \ \nu>0,
\end{array}
\label{eq:SPDEformulaWeakSolution}
\end{equation}
where $\mathbf{s}$ are the locations; $\kappa$ is a spatial scale parameter positive defined and related to $\rho$ and $\nu$ through $\kappa=\sqrt{8\nu}/\rho$; $\alpha$ controls the smoothness of the performances, which by default is $2$; $\tau$ regulates the variance, so it is also a parameter defined positive as $\tau^2=\Gamma(\nu)\left[ \Gamma(\alpha)(4\pi)^{d/2}\kappa^2\sigma^2\right]^{-1}$; and $\mathcal{D}$ is the spatial domain $\mathcal{D}\subset \mathbb{R}^d$, being $d$ the Euclidean spatial dimension of such spatial domain. Additionally, $\Delta=\sum_i\partial^2/\partial s_i^2$ is the Laplacian operator and $\mathcal{W}$ denotes a spatial stochastic Gaussian process with unit variance. Therefore, the solution for $\xi(\mathbf{s}_i)$ leads to a Gaussian field with a covariance matrix defined by the Matérn function correlation $\mathcal{C}(h)$, where the covariance is the correlation multiplied by the marginal deviation:
\begin{equation}
    \mathcal{C}(h)=\sigma^2\cdot\frac{2^{1-\nu}}{\Gamma(\nu)}\left(\sqrt{2\nu}\cdot h /\rho\right)\mathcal{K}_\nu\left(\sqrt{2\nu}\cdot h/\rho\right).
\end{equation}
In the above equation, $\sigma$ is the marginal standard deviation, $\rho$ is the spatial range, $\mathcal{K}_\nu$ is a modified Bessel function of the second kind of order $\nu$, where $\nu$ is a parameter of smoothness defined as $\nu=\alpha-d/2$. In INLA, $\alpha=2$ is taken by default, and in our cases, the dimension of the analysis space is the plane ($d=2$), resulting in a constant smoothness parameter value $\nu=1$. Furthermore, under the SPDE approximation, the spatial range $\rho$ is considered as the distance at which the correlation value is close to $0.1$ \citep{Lindgren_ExplicitLinkSPDE_2011}. 

This approach allows defining a spatial effect whose precision structure is sparse and approximates the solution of a continuous Gaussian field with a variance-covariance structure belonging to the Matérn family; where one particular case of this family is the exponential covariance function when $\nu = 1/2$. This spatial effect is defined by a Gaussian Markov Random Field (GMRF), $\mathbf{u} \sim \text{GMRF}(\mathbf{0}, \mathbf{Q})$, whose precision matrix for the standard case of $d=2$ and $\nu = 1$ can be expressed as $\mathbf{Q} = \tau^2 (\kappa^4 \mathbf{C} + 2\kappa^2 \mathbf{G} + \mathbf{G} \mathbf{C}^{-1}\mathbf{G})$. Here, $\mathbf{C}$ and $\mathbf{G}$ are two matrices that can be computed through the structure of a constrained refined Delaunay triangulation $\mathcal{T}$ by Finite Element Methods. The GMRF realisations at the node locations $\mathbf{s}_n$ can be projected to any other point within the mesh domain, usually to the observation locations $\mathbf{s}_o$, through a linear basis approximation synthesised in the matrix product of the projection matrix $\mathbf{A}(\mathbf{s}_n, \mathbf{s}_o)$ with the GMRF realisation at the mesh nodes $\mathbf{u}(\mathbf{s}_{n})$, $\mathbf{u}(\mathbf{s}_{o})=\mathbf{A}(\mathbf{s}_n, \mathbf{s}_o)\mathbf{u}(\mathbf{s}_{n})$. This allows the spatial effect $\mathbf{u}(\mathbf{s}_o)$ to be rewritten as $\mathbf{u}(\mathbf{s}_{o}) \sim \text{GMRF}(\mathbf{0}, \mathbf{A}^T\mathbf{Q}\mathbf{A})$, which is clearly equivalent to $\mathbf{u}(\mathbf{s}_{o}) = \mathbf{A}(\mathbf{s}_n, \mathbf{s}_o) \mathbf{u}(\mathbf{s}_{n})$.

\subsection{Downscaling model with SPDE-FEM}

The definition of a spatial effect whose structure is continuous in space and space-time allows for spatial and spatio-temporal downscaling. Let's assume we have the following structure for the linear predictors of the model of our data: 
\begin{equation}
\begin{array}{rcl}
     \eta_{i,1} & = & \mathbf{X}\boldsymbol\beta_1 + \sum_{k=1}^Kf_{k,1}(z_{i}) + u_{st,1}(C_i,t_i)  \\
     & \vdots & \\
     \eta_{i,D} & = & \mathbf{X}\boldsymbol\beta_D + \sum_{k=1}^Kf_{k,d}(z_{i}) + u_{st,D}(C_i,t_i)
\end{array}
\label{eq:linear_predictors_downscaling}
\end{equation}
the downscaling is defined for the spatial structure when, for each area $C_i$, the spatially structured effect is calculated as follows:
\begin{equation}
    u_{st}(C_i, t_i) = \frac{\int_{\mathbf{s} \in C_i} u_{st}(\mathbf{s},t_i) \text{d}\mathbf{s}}{|C_i|} \approx \frac{\sum_{i=1}^{n(\mathbf{s}_i\in C_i)} u_{st}(\mathbf{s}_i,t_i)}{n(\mathbf{s}_i\in C_i)} = \frac{\mathbf{A}(\mathbf{s}_n, s_i)\mathbf{u}_{st}(\mathbf{s}_n,t_i)}{n(\mathbf{s}_i\in C_i)}
\end{equation}
where $s_i$ are integration points, allowing the approximation of the integral in the region $C_i$ by averaging the sum of the field at the integration points, and $n(s_i \in C_i)$ is the number of integration points in $C_i$; being the area of $C_i$ determine as $|C_i| = \int_{\mathbf{s} \in C_i} \text{d}\mathbf{s}$. If a spatio-temporal downscaling is implemented, then the integration would be performed in both dimensions, spatial $s$ and temporal $t$, using integration points for each corresponding volume $C_i \times t_i$. Additionally, this procedure can also be applied to establish sub-models for those covariates with a spatial structure that are given in areas. These sub-models would involve the modelling of such covariates by incorporating the integral of the covariate scaled by a linear regression coefficient, analogous to how it is done in error models \citep{Muff_ErroModel_2015, Muff_TwoComponentErroModel_2017} in cases without downscaling. The linear predictors in Equation (\ref{eq:linear_predictors_downscaling}) can be associated with either a Dirichlet distribution or a multivariate normal distribution related to some of the logratio transformations previously shown. If they correspond to a logratio transformation, it is necessary to re-write the linear predictors to incorporate the variance-covariance structure $\boldsymbol\Sigma$ of the multivariate normal distribution, as explained in previous sections.

The downscaling approach not only provides a way to obtain estimates on a continuous scale of the spatial effect but, due to its structure and implementation, also allows us to consider the same spatial effect for different spatial supports. This has several implications, as detailed in the initial references. One particular implication we would like to highlight is that when the spatial support varies over time—such as in the case of small areas for disease mapping or NUTS3—the use of a downscaled spatial or spatio-temporal effect ensures a coherent structure across the different supports, maintaining the spatial or spatio-temporal dependence of the data. Furthermore, this does not depend on the specific use of the SPDE-FEM approach, as it can be implemented with any other Gaussian field, such as spline and two-dimensional spline bases, or any other continuous field.

\section{Big Data}


Public smart card data, Wi-Fi access point data, wireless sensor networks, and data with spatial location information —such as social media data, mobile phone tracking, and other sensing information from Internet of Things devices— can provide useful ancillary data for LULC (Land Use and Land Cover) mapping. Compared to traditional geospatial data acquisition, these geospatial big data (GBD) are typically obtained at a lower cost and offer different coverages and better spatio-temporal resolutions \citep{Liu_BigDataLandUse_2020, Zhang_LandUseBigData_2022}. They contain abundant human activity information, which can compensate for the lack of socioeconomic data \citep{Liu_BigDataLandUse_2020}. By leveraging both Remote Sensing (RS) and GBD, it is possible to examine the physical and socioeconomic characteristics of the urban land system \citep{Marti_UrbanStudiesSocialMedia_2019}. Despite the great potential of integrating RS and GBD for enhanced insights into urban land use, significant challenges remain in storing, managing, analysing, and visualising these data due to differences in spatial data quality (e.g., semantics, timestamps, and scale), technical formats, and data structures \citep{Li_GeospatialBigData_2016}. Therefore, a large amount of land use information is accessible, particularly with global and low-scale land use data, which gives rise to a big data problem. 

{
\linespread{1.}
\begin{figure}
    \centering
    \begin{tikzpicture}
    \node at  (0,0) [rectangle,draw] (1) {$\everymath={\displaystyle}
    \begin{array}{c}
    \text{\underline{Modelling}} \; \mathbf{y}_1 \\[0.2cm]
    \pi(\boldsymbol\beta\mid\mathbf{y}_1)\\[0.2cm]
    \pi(\mathbf{x}_{-\boldsymbol\beta}\mid\mathbf{y}_1)\\[0.2cm]
    \pi(\boldsymbol\theta\mid\mathbf{y}_1)\\
    \end{array}$};

    \node at  (2.25,0) [rectangle] (prior_to_2) {$\everymath={\displaystyle}
    \begin{array}{c}
    \pi(\boldsymbol\beta\mid\mathbf{y}_1)\\[0.2cm]
    \pi(\boldsymbol\theta\mid\mathbf{y}_1)\\
    \end{array}$};
    
    \node at  (4.5,0) [rectangle,draw] (2) {$\everymath={\displaystyle}
    \begin{array}{c}
    \text{\underline{Modelling}} \; \mathbf{y}_2 \\[0.2cm]
    \pi(\boldsymbol\beta\mid\cup_{i=1}^2\mathbf{y}_i)\\[0.2cm]
    \pi(\mathbf{x}_{-\boldsymbol\beta}\mid\mathbf{y}_2)\\[0.2cm]
    \pi(\boldsymbol\theta\mid\cup_{i=1}^2\mathbf{y}_i)\\
    \end{array}$};

    \node at  (7.1,0) [rectangle] (prior_to_n) {$\everymath={\displaystyle}
    \begin{array}{c}
    \ldots\pi(\boldsymbol\beta\mid\cup_{i=1}^{n-1}\mathbf{y}_i)\\[0.2cm]
    \ldots\pi(\boldsymbol\theta\mid\cup_{i=1}^{n-1}\mathbf{y}_i)\\
    \end{array}$};

    \node at  (9.8,0) [rectangle,draw] (n) {$\everymath={\displaystyle}
    \begin{array}{c}
    \text{\underline{Modelling}} \; \mathbf{y}_n \\[0.2cm]
    \pi(\boldsymbol\beta\mid\cup_{i=1}^n\mathbf{y}_i)\\[0.2cm]
    \pi(\mathbf{x}_{-\boldsymbol\beta}\mid\mathbf{y}_n)\\[0.2cm]
    \pi(\boldsymbol\theta\mid\cup_{i=1}^n\mathbf{y}_i)\\
    \end{array}$};

    \node at  (3,-4.5) [rectangle,draw] (x) {$\everymath={\displaystyle}
    \begin{array}{c}
    \text{\underline{Combining random effect posteriors}} \\[0.2cm]
    1.\;\text{Marginals}\quad x_i \approx \sum_{j=1}^n w_{ij}x_{ij}  \\[0.2cm]
    2.\;\text{Multivariate}\quad \pi(\mathbf{x}_i\mid \mathbf{y})\approx \prod_{j=1}^n \pi(\mathbf{x}_i\mid \mathbf{y}_j)
    \end{array}$};

    \node at  (10.5,-2.25) [rectangle] (posterior_n) {$\everymath={\displaystyle}
    \begin{array}{c}
    \pi(\boldsymbol\beta\mid\mathbf{y})\\[0.2cm]
    \pi(\boldsymbol\theta\mid\mathbf{y})\\
    \end{array}$};    

    \node at  (9.8,-4.5) [rectangle,draw] (result) {$\everymath={\displaystyle}
    \begin{array}{c}
    \text{\underline{Posterior distributions}} \\[0.2cm]
    \pi(\boldsymbol\beta\mid\mathbf{y})\\[0.2cm]
    \pi(\mathbf{x}_{-\boldsymbol\beta}\mid\mathbf{y})\\[0.2cm]
    \pi(\boldsymbol\theta\mid\mathbf{y})\\
    \end{array}$};

    \draw[-to] (1) to [out=0,in=180] (2);
    \draw[-to] (2) to [out=0,in=180] (n);

    \node at  (4.5,2.3) [rectangle] (alg2) {$\everymath={\displaystyle}
    \begin{array}{c}
    \pi(\boldsymbol\beta\mid\mathbf{y}), \; \pi(\boldsymbol\theta\mid\mathbf{y})
    \end{array}$};

    \draw[dotted,-to] (n) to [out=90,in=90,looseness=0.25] (1);
    \draw[dotted,-to] (n) to [out=90,in=90,looseness=0.25] (2);

    \draw[double,-to] (1) to [out=-90,in=90,looseness=0.95] (x);
    \draw[double,-to] (2) to [out=-90,in=90,looseness=1.25] (x);
    \draw[double,-to] (n) to [out=-90,in=90,looseness=0.75] (x);

    \draw[-to] (n) to [out=-90,in=90,looseness=1.5] (result);
    \draw[double,-to] (x) to [out=0,in=180,looseness=0.5] (result);
    \end{tikzpicture}
    \vspace{5mm}
    \caption{Scheme of the sequential consensus procedure. This approach assumes sequentially updating the fixed effects and hyperparameters, followed by performing a consensus for the random effects after these sequential updates \citep{Figueira_SequentialConsensus_2024}.}
    \label{fig:Scheme_sequentialconsensus}
\end{figure}
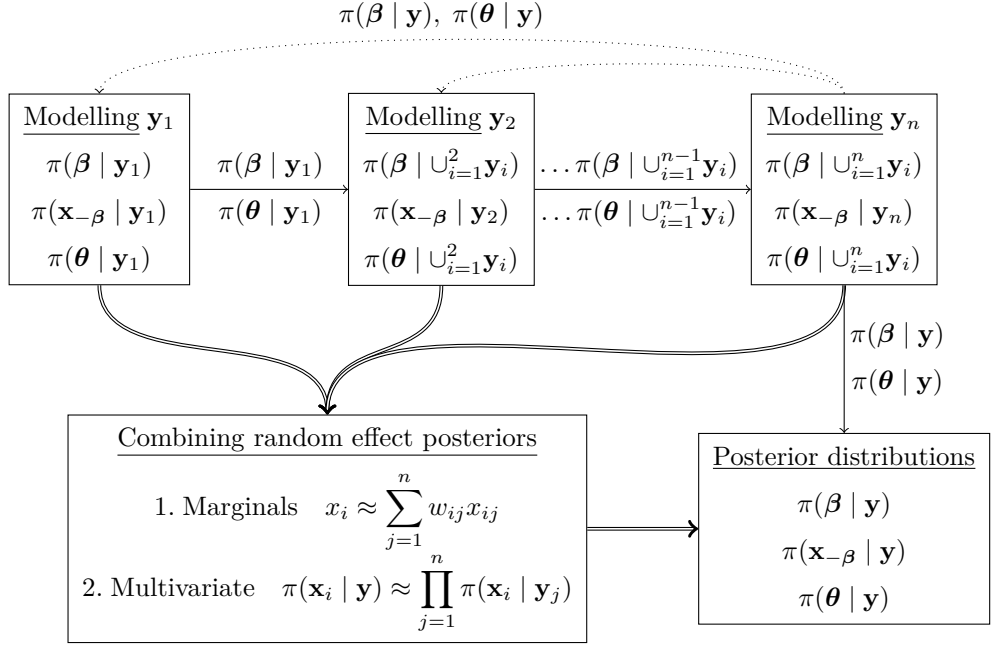
}

This entails dealing with very heterogeneous and high dimensional data, particularly when data are available at global scale \citep{Stanimirova_GlobalLandUse_2023} or when LULC data at regional scales are available at fine grid \citep{You_DisaggregationEntropy_2005, Chakir_SpatialDownscaling_2009}. The joint analysis of data that are at different scales and whose spatial structure is in different supports, as well as combining different sources of information, means that modelling structures can be particularly complex \citep{Figueira_BayesianFeedback_2023, Figueira_SequentialConsensus_2024}. In order to deal with these large databases and complex modelling structures, sequential inference procedures can be implemented in INLA, which reduces the computational cost. In particular, a sequential consensus inferential procedure can be established and implemented in \texttt{R-INLA} \citep{Figueira_SequentialConsensus_2024}. This procedure, summarised in Figure \ref{fig:Scheme_sequentialconsensus}, involves the marginal updating of fixed effects and hyperparameters over a given partition of the data. The partition of the data can be done along the various sources of information available, such as different likelihoods or groups of likelihoods, or by leveraging the structure of the latent field (spatial, temporal, or spatio-temporal). Once the sequential inference is performed, the random effects information is combined using a consensus approach according to their (i) marginal or (ii) multivariate distributions. This approach also accommodates complex models with shared components between different likelihoods.

This procedure reduces computational burden in terms of memory and CPU usage by solving the dataset or the original model through partitions. While it provides good estimates for the latent field, the evaluation of hyperparameters may differ from that obtained by analysing the full model without partitioning. This is because only the marginal information is updated, and hyperparameters can exhibit non-negligible correlations in the posterior distribution.

\section{Examples}

In this section, several examples are provided to illustrate the implementation of the various methodological approaches described in the previous sections. These range from handling zeros and ones using adaptations of Hurdle models—widely used in environmental and ecological sciences—to methods for downscaling and procedures for managing large databases or complex models.

The first example demonstrates how to handle compositional data with 0's and 1's, using simulated data. The second example illustrates the results of a downscaling model applied to real land use data from the European LAMASUS project. Finally, the third example uses a large simulated dataset to showcase the effectiveness of the sequential consensus procedure in analysing extensive datasets and accurately estimating the underlying process.

\subsection{Dealing with 0's example}

In this first example, we present two cases that illustrate how to handle the presence of zeros in CoDa. The first case uses data simulated from a Beta distribution to exemplify the simplest scenario, proving the Hurdle model for dealing with zeros. The second case also addresses CoDa, analysing $3$ categories through a multivariate model of the CLR transformation.

\begin{figure}
    \centering
    \includegraphics[width=\linewidth]{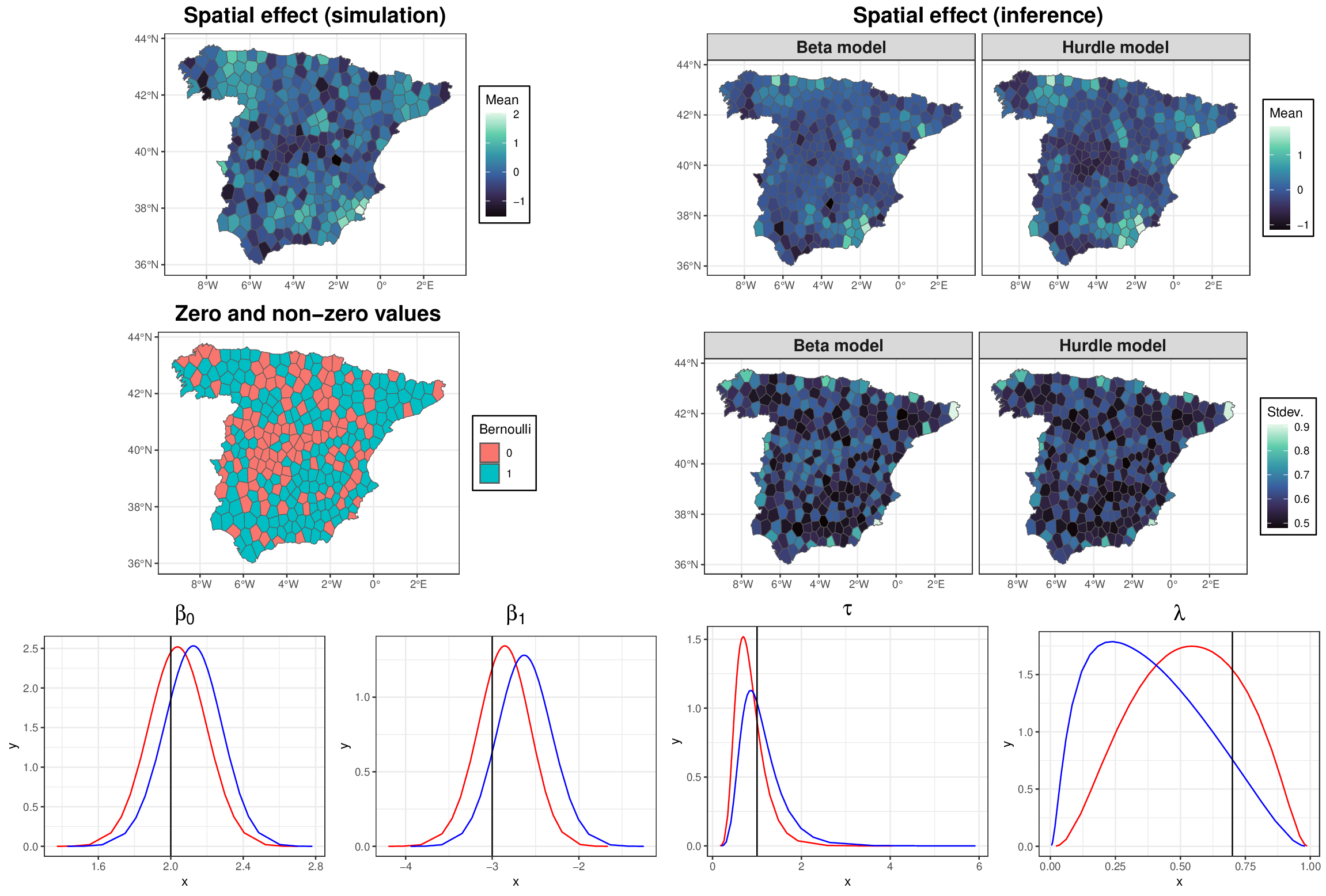}
    \caption{Simulated spatial effect, spatial distribution of the zero and non-zero values, and inferred spatial effect (mean and stdev.), along with the posterior distributions of the fixed parameters and hyperparameters. Distributions from the Hurdle model are shown in red, and those obtained by the Beta model are shown in blue.}
    \label{fig:BetaZeros_example}
\end{figure}

In the Beta case, we simulate the data from the following hierarchical conditional model:
\begin{equation}
\begin{array}{rcl}
      Z_i & \sim & \text{Ber}(\pi_i), \\
      \text{logit}(\pi_i) & = & \beta_{B} + \alpha\cdot(\beta_0 + \beta_1 \cdot x_i + u_i), \\
      Y_j \mid Z_j = 1 & \sim & \text{Beta}(\mu_j, \phi), \\
      \text{logit}(\mu_j) & = & \beta_0 + \beta_1 \cdot x_j + u_{sj}, 
\end{array}
\end{equation}
where $Z_i$ are Bernoulli random variables explaining the process that generates the incidence vector (incidence matrix for CoDa), such that $z_i=0 \iff y_i=0$ and $z_1 \iff y_i \neq 0$. $Y_j$ are Beta random variables conditioned on $Z_j=1$, which implies that $Y_j\neq 0$. The latent field comprises an intercept $\beta_0$, a linear coefficient $\beta_1$ related to a covariate $\mathbf{x}$, and a spatial effect $\mathbf{u}_s$ following a Leroux structure. The spatial adjacency structure is built by partitioning mainland Spain with a Voronoi diagram, as shown in Figure \ref{fig:BetaZeros_example}. This latent field is shared in the linear predictor of the Bernoulli process, scaled by $\alpha$ and including an intercept $\beta_B$ that controls the overall proportion of 0's. Finally, $\phi$ stands for the precision of the Beta distribution.

Figure \ref{fig:BetaZeros_example} shows the simulated spatial structure and the distribution of null and non-zero values, along with the inferential results of the beta model without incorporating the zeros and the Hurdle model incorporating the null values. It can be seen that, in general, the Hurdle model better captures the true values of the spatial effects where there are many zeros, as well as the posterior distributions of the model parameters.

The CoDa simulation is performed considering that only one ($\mathbf{Y}_1$) of the $3$ categories ($\mathbf{Y}_{1}, \mathbf{Y}_{2}, \mathbf{Y}_{3}$) can present zero values. This model serves as a conceptual example of how it can be implemented for CoDa, dealing with zeros and ones in several categories. Therefore, the model structure when no null values are present is as follows:
\begin{equation}
\begin{array}{c}
     Z_{i1} \sim \text{Ber}(\pi_i),  \\
     Y_{j1} \mid Z_{j1} = 1 \sim \text{N}(\mu_{j1}, \tau^*), \\
     Y_{j2} \mid Z_{j1} = 1 \sim \text{N}(\mu_{j2}, \tau^*), \\
     Y_{j3} \mid Z_{j1} = 1 \sim \text{N}(\mu_{j3}, \tau^*), \\
\end{array}
\end{equation}
where the linear predictor for the CLRs are defined with the same structure as the one used for the Beta example with the addition of the $\mathbf{u}$ effect to take into account the correlation between the categories, $\eta_{id} = \beta_{0d} + \beta_{1d} x_{i} + u_{sid} + u_{i}$. However, when $Z_{i1} = 0$ the CLR change to the remaining two categories ($\mathbf{Y}_2, \mathbf{Y}_3$) using the same underlying process in the linear predictors as before.

\begin{figure}
    \centering
    \includegraphics[width=\linewidth]{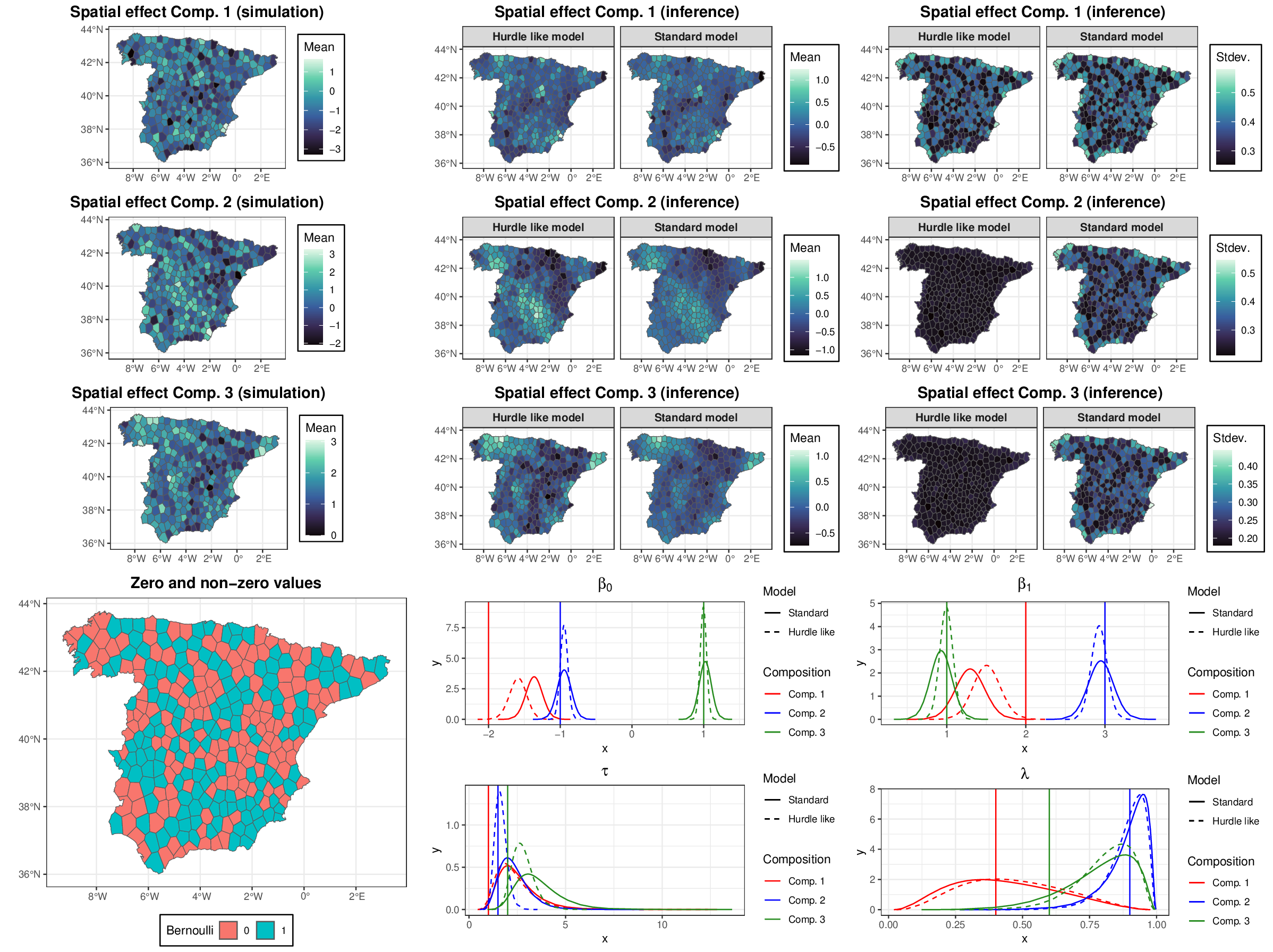}
    \caption{Simulated spatial effects, spatial distribution of the zero and non-zero values, and inferred spatial effects (mean and stdev.), along with the posterior distributions of the fixed parameters and hyperparameters.}
    \label{fig:CoDaZeros_example}
\end{figure}

Figure \ref{fig:CoDaZeros_example} shows the simulated spatial effects and the distribution of the zero and non-zero values for the first composition (\textit{Comp. 1}). The figure also shows the inferred spatial effects (mean and stdev.) for the different compositions, along with the distributions for the fixed parameters and hyperparameters. 

In both cases we can observe an improvement in the estimation of the model by incorporating the Hurdle-type structure to integrate the analysis of the zeros.

\subsection{Downscaling example: LAMASUS data}


In this example, we present the implementation of the downscaling procedure for real land use data from the European LAMASUS project.  In the context of the LAMASUS project, we aim to identify the components that determine or drive land use and land use change in Europe over the last decades. The data used in this example is spatially structured at the NUTS3 level between the years 2007 and 2018; where NUTS (Nomenclature of Territorial Units for Statistics) is the hierarchical system for dividing the economic territory of the European Union. The land use data analysed in this example is classified into five aggregated categories to avoid the presence of zero and unit values. These categories are: (i) cropland, (ii) grassland, (iii) forest, (iv) urban, and (v) other natural land. A wide range of explanatory variables are available to analyse the underlying processes that determine land use, including socio-economic and demographic variables such as GDP, GVA, and employment, as well as environmental variables like aspect and elevation.  


We begin by focusing on the proportion of agricultural land use, specifically the cropland category. To model exclusively the proportion of cultivated land use, we employ a downscaling model using a Beta distribution. Specifically, we use a spatio-temporal hierarchical Bayesian model, as outlined below:
\begin{equation}
\begin{array}{rcl}
    Y_{it} & \sim & \text{ Beta}\left(\mu_{it}, \phi \right) \\  \\
    \text{logit}\left(\mu_{it}\right) & = & \boldsymbol{\beta} \mathbf{}{X}_{it} + \int_{\mathbf{s} \in C_{i}} \frac{u(\boldsymbol{s}) \text{d}\mathbf{s}}{\mid C_i \mid} + u_t, \\
\end{array}
\end{equation}
where $Y_{it}$ represents the random variable for the proportion of cropland use in a specific region $i$ and year $t$. The conditional distribution of $Y_{it}$ follows a Beta distribution with mean $\mu_{it}$ and precision $\phi$. The mean is linked through the logit function to the linear predictor. The covariates are represented by $\mathbf{X}_{it}$, and the vector of the linear regression coefficients are $\boldsymbol\beta$. The downscaling spatial effect is $u(\mathbf{s})$, while the temporal effect is modelled as a first-order autoregressive process $u_t=\phi_t u_{t-1} + \epsilon(\tau_t)$, where $\phi_t$ and $\tau_t$ are the temporal autoregressive parameter and the marginal precision of the temporal autoregressive effect, respectively. The selection of covariates to be included in the model was performed by exploring all possible combinations of models without spatial and temporal effects, selecting the combination with lower Watanabe-Akaike Information Criterion (WAIC, \citealt{Watanabe_WAIC_2013}) is selected. 

\begin{figure}
    \centering
    \includegraphics[width=\linewidth]{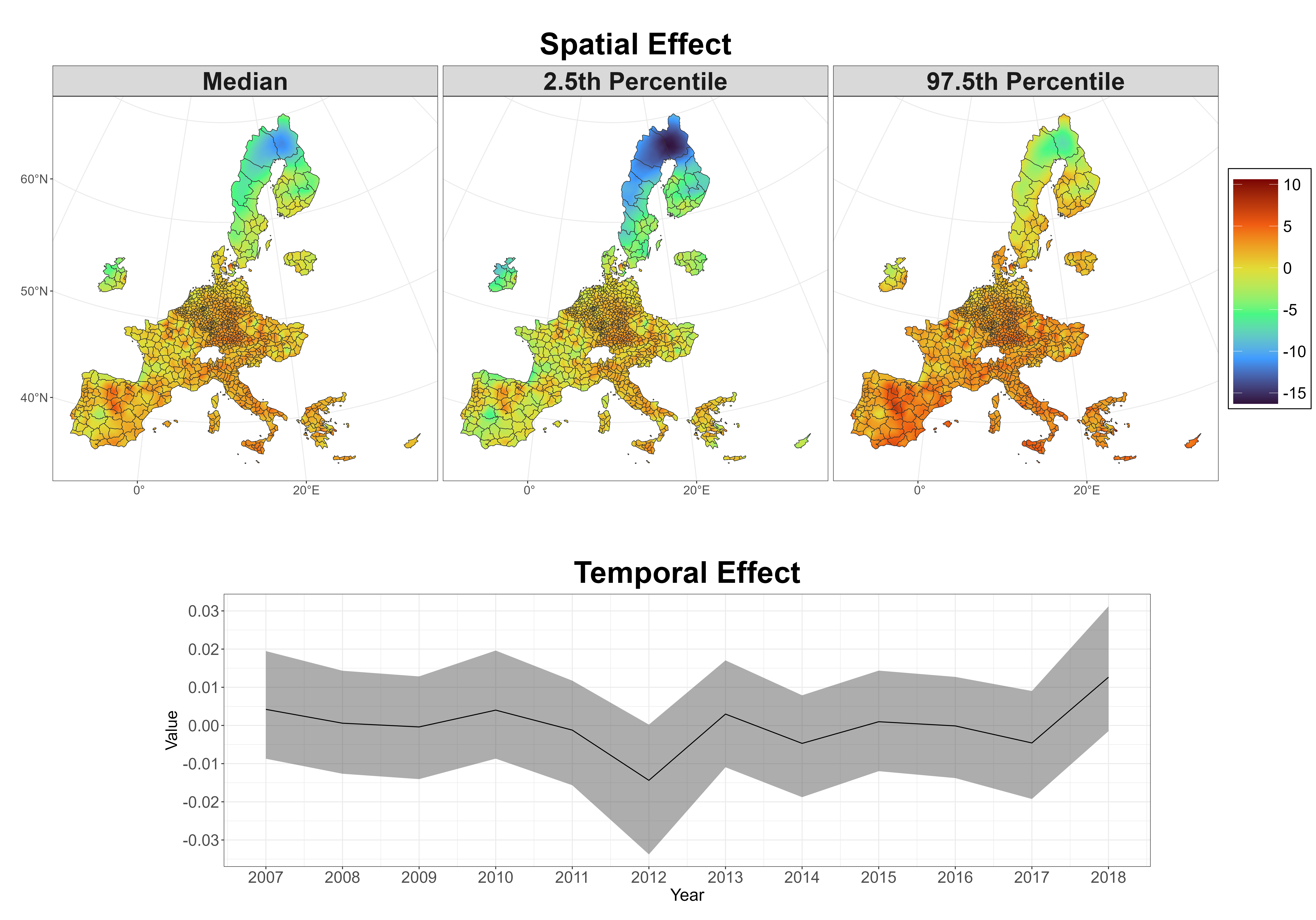}
    \caption{Cartographic representation of the continuous spatial effect and the temporal effect of the downscaling model.}
    \label{fig:fig_BetaExample}
\end{figure}

Some results can be seen in Figure \ref{fig:fig_BetaExample}. As the figure shows, the temporal effect is small. As for the spatial effect, it captures the variability not explained by the covariates. We can observe that, in certain areas, the spatial effect has a greater impact on the overall average effect than in other regions, where its impact is less pronounced. Regarding the covariates, some notable ones include decoupled payments (in euros) normalized by the total used agricultural area (in hectares), and less favoured area subsidies (in euros) normalized by the total used agricultural area (in hectares). Additionally, other relevant variables are the area of the region, elevation of the terrain, GVA, and total output (in euros).

The analysis for the $5$ aggregated land use categories is performed using a multivariate Gaussian model with the ALR transformation of the categories. To select the covariates, we implemented a stepwise search algorithm driven by the WAIC of the models after a pre-selection where high correlated ($\rho>0.75$) variables were extracted. This algorithm involves the following steps: (i) a forward's search in the covariates until no new covariate can be included, then (ii) a backward step until no covariate can be deleted, and (iii) repeating steps (i) and (ii) until there is no change. The base model into which the algorithm is implemented encompasses a specific intercept, spatial, and temporal component for each ALR. The final model includes $9$ explanatory covariates, and the model for the mean $\boldsymbol\mu = (\boldsymbol\mu_1, \ldots, \boldsymbol\mu_4)$ of the multivariate Gaussian distribution $\text{MVN}(\boldsymbol\mu, \boldsymbol\Sigma)$ is written as follows:
\begin{equation}
\begin{array}{rcl}
     \mu_{i1} & = & \beta_{01} + \mathbf{X}_{i}\boldsymbol\beta_{1} + u_{si1} + u_{ti1} + u_i,\\
     \mu_{i2} & = & \beta_{02} + \mathbf{X}_{i}\boldsymbol\beta_{2} + u_{si2} + u_{ti2} + u_i,\\
     \mu_{i3} & = & \beta_{03} + \mathbf{X}_{i}\boldsymbol\beta_{3} + u_{si3} + u_{ti3} + u_i,\\
     \mu_{i4} & = & \beta_{04} + \mathbf{X}_{i}\boldsymbol\beta_{4} + u_{si4} + u_{ti4} + u_i,\\
\end{array}
\end{equation}
where $u_{sid}$ is the downscaled spatial effect, $u_{sid} = \int_{\mathbf{s}\in C_i} u_{sd}(\mathbf{s})\text{d}\mathbf{s} /|C_i|$, and the component $\mathbf{u}$ accounts for the correlation $\boldsymbol\Sigma^{-1}$ between the categories, as explained in the corresponding section.

\begin{figure}
    \centering
    \includegraphics[width=\linewidth]{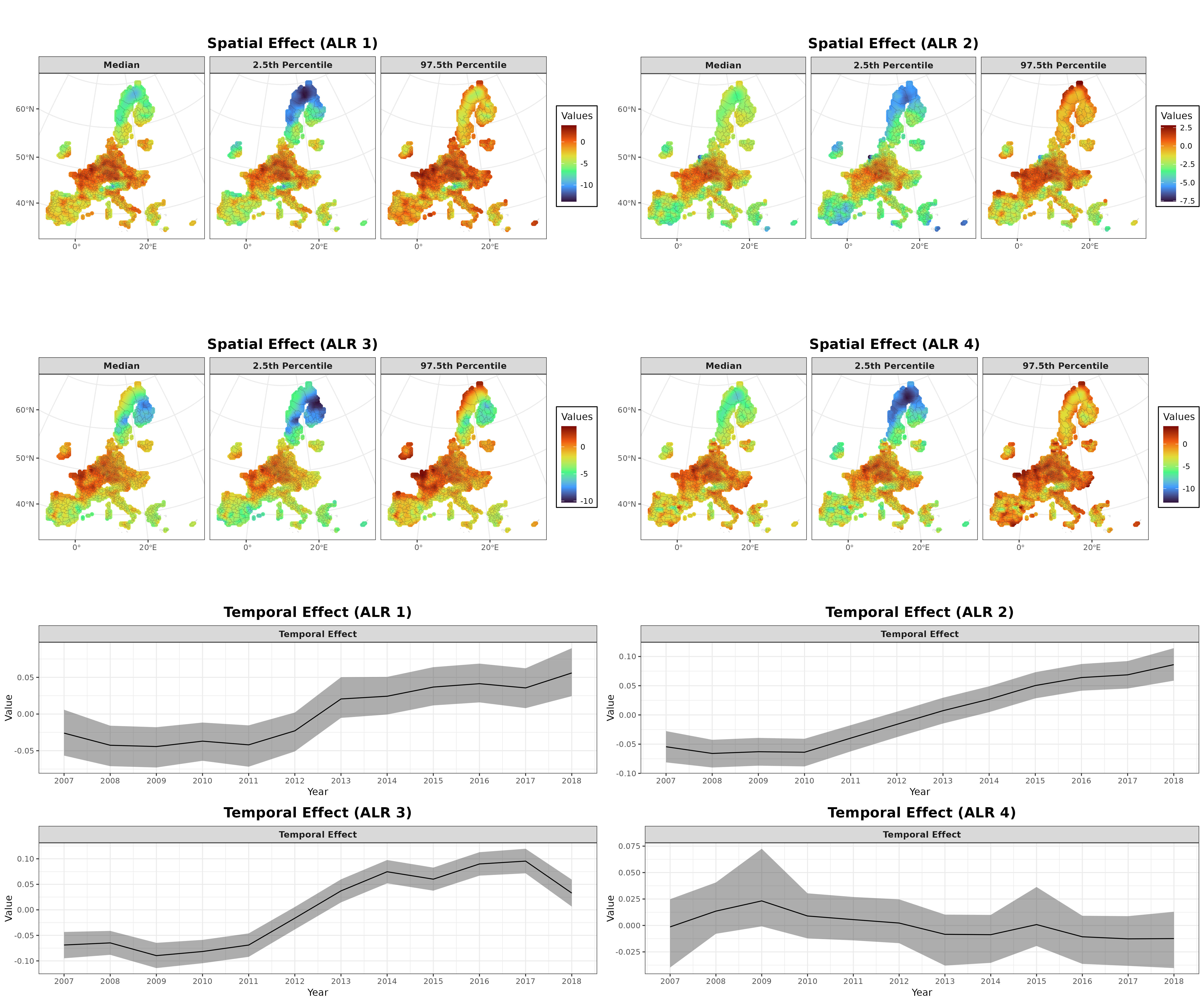}
    \caption{Cartographic representation of the continuous spatial effects and the temporal effects for the different ALRs of the downscaling model.}
    \label{fig:fig_CoDaExample}
\end{figure}

Figure \ref{fig:fig_CoDaExample} shows the downscaled continuous spatial effects for the different ALRs, together with the temporal effects. In this case, we can observe the scaling effect for the different ALRs, which implies a common spatial dependency structure over the years. This is combined with the temporal effect for the log-ratios, where the trend is similar for the first three, while the fourth shows a spatial effect with greater uncertainty and is less pronounced.

\subsection{Big Data example}

In this final example, we present the implementation of the algorithm summarised in Figure \ref{fig:Scheme_sequentialconsensus} for evaluating large compositional data databases. The purpose of the simulation is not merely to demonstrate the results of the algorithm implementation for handling large databases, but also to showcase its versatility and its application to spatio-temporal downscaling models. This enables the processing of structured data in areas with spatial support that varies over the years.

The simulation is conducted for three categories evaluated using the ALR transformation, so the multivariate Gaussian structure will have a dimension of $2$. The spatial structure is simulated using a continuous Gaussian field aggregated over two different spatial supports structured in areas, making the downscaling model purely spatial. The Gaussian field is a separable spatio-temporal effect, defined by the following precision matrix: $\mathbf{Q}_{st} = \mathbf{Q}_s \otimes \mathbf{Q}_t$. The spatial precision matrix is defined using an SPDE in two dimensions, while the temporal part is determined by a first-order autoregressive structure with $600$ (temporal) nodes. To construct the different spatial supports, the territory of mainland Spain has been partitioned using Voronoi diagrams.



The model used to simulate and infer the data is as follows:
\begin{equation}
\begin{array}{rcl}
\mu_{it1} = \beta_{01} + \int_{\mathbf{s}\in C_i}u_1(\mathbf{s},t)\text{d}\mathbf{s} + u_{it}, \\
\mu_{it2} = \beta_{02} + \int_{\mathbf{s}\in C_i}u_2(\mathbf{s},t)\text{d}\mathbf{s} + u_{it}, \\
\end{array}
\end{equation}
where $\beta_{0d}$ are the intercepts, $\mathbf{u}_d(\mathbf{s},t)$ are the spatio-temporal effects aggregated over the different spatial supports for each ALR. Finally, the effect $\mathbf{u}$ accounts for the correlation structure between the different ALRs, allowing for the implementation of the multivariate Gaussian distribution in \texttt{R-INLA}.

\begin{figure}
    \centering
    \includegraphics[width=\linewidth]{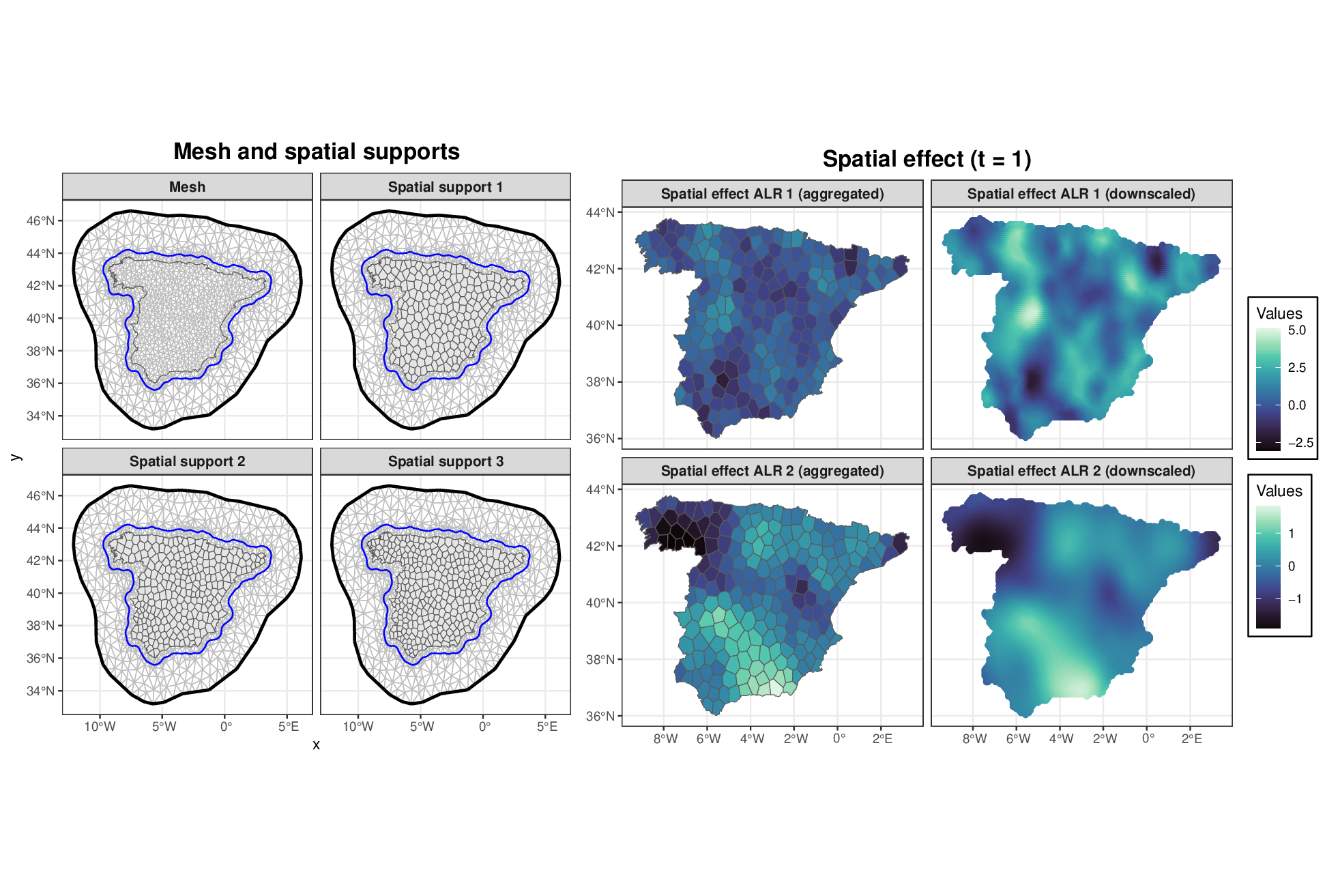}
    \caption{Mesh for spatial effects, spatial supports for data aggregation and outputs, comparing the aggregated simulated data with the downscaled spatial fields obtained from the inference for the first temporal node.}
    \label{fig:fig_BigDataExample}
\end{figure}

Figure \ref{fig:fig_BigDataExample} shows the mesh for the spatial effect and the different spatial supports in which the data was aggregated over the various years. It also presents the aggregated spatial effect using simulated data, as well as the desegregated spatial effects obtained from the inferential process of the two ALRs for the first temporal node. The data set  was divided along the temporal scale, grouping into 6 temporal nodes, as larger grouping would exceed the server's memory capacity, which had 253 GB of RAM and 93 cores. Additionally, the complete analysis by the sequential consensus procedure for the entire dataset took $314.28$ minutes. 

\section{Conclusions}

In the field of land use science, the application of spatial or spatio-temporal models is essential for evaluating the key factors that modify and direct land use and land use changes. Assessing spatial patterns alongside temporal patterns provides a precise understanding of their distribution in space and their temporal evolution. To achieve this, approaches often use either large-scale aggregated data or small-scale disaggregated data. In the latter case, when small-scale data is available, two natural problems may arise: the presence of null values and computational complications when handling large databases. Conversely, when dealing with large-scale data, the issue of changing the spatial aggregation support over time may occur. This study presents methodologies to address the problem of null values, downscaling procedures for spatial or spatio-temporal structures, and big data challenges. These methodologies are illustrated with various examples to demonstrate their implementation.

The results presented in the examples show the improvement when the presence of zeros is integrated into the analysis, along with the ability to perform dowscaling or disaggregation models using continuous Gaussian fields, specifically through the SPDE-FEM approach. Additionally, the implementation of a dowscaling model on a large dataset across different aggregation supports was shown. This highlights the possibility of encountering aggregated data in different spatial structures over the years, and how to deal with them in a Big Data context. This can be achieved by implementing the sequential consensus algorithm; otherwise, computational limitations would render the analysis unfeasible. 



\section*{Acknowledgments}


DC and ALQ thank support by the grant PID2022-136455NB-I00, funded by Ministerio de Ciencia, Innovación y Universidades of Spain (MCIN/AEI/10.13039/501100011033/FEDER, UE) and the European Regional Development Fund. DC also acknowledges Grant CIAICO/2022/165 funded by Generalitat Valenciana. 


This study is part of the ThinkInAzul program and is financed by the MCIN with funds from the European Union (NextGenerationEU-PRTR-C1711) and by the Generalitat Valenciana GVA-THINKINAZUL/2021/021. 

\section*{Conflict of Interest}

The authors declare no conflict of interest.

\bibliography{bibliography}

\end{document}